\documentclass[showpacs,twocolumn,floatfix,superscriptaddress]{revtex4-1}

\usepackage{graphicx}     
\usepackage{dcolumn}      
\usepackage{amsmath}    
\usepackage{amssymb}      
\usepackage{bm}           

\begin{document}

\title{ Metallic clusters on a model surface: quantum versus geometric effects }

\author{S. A. Blundell}
\affiliation{SPSMS, UMR-E CEA/UJF-Grenoble 1, INAC, Grenoble, F-38054, France.}

\author{Soumyajyoti Haldar}
\email[Corresponding author: ]{soumyajyoti.haldar@gmail.com}
\altaffiliation{\\Current Affiliation: Department of Physics and Astronomy, Uppsala
University, Box 516, 75120 Uppsala, Sweden}
\affiliation{Centre for Modeling and Simulation, University of Pune, Ganeshkhind, Pune
411 007, India}
\affiliation{Department of Physics, University of Pune, Ganeshkhind, Pune 411 007, India.}

\author{D. G. Kanhere}
\affiliation{Department of Physics, University of Pune, Ganeshkhind, Pune 411 007, India.}


\begin{abstract}

We determine the structure and melting behavior of supported metallic clusters using an
\textit{ab initio} density-functional-based treatment of intracluster interactions and an
approximate treatment of the surface as an idealized smooth plane yielding an effective
Lennard-Jones interaction with the ions of the cluster.  We apply this model to determine the structure of sodium clusters containing from 4 to 22 atoms, treating the cluster-surface interaction strength as a variable parameter.  For a strong cluster-surface interaction, the clusters form two-dimensional (2D) monolayer structures; comparisons with calculations of structure and dissociation energy performed with a classical Gupta interatomic potential show clearly the role of quantum shell effects in the metallic binding in this case, and evidence is presented that these shell effects correspond to those for a confined 2D electron gas.  The thermodynamics and melting behavior of a supported Na$_{20}$ cluster is considered in detail using the model for several cluster-surface interaction strengths. We find quantitative differences in the melting temperatures and caloric curve from density-functional and Gupta treatments of the valence electrons.  A clear dimensional effect on the melting behavior is also demonstrated, with 2D structures showing melting temperatures above those of the bulk or (at very strong cluster-surface interactions) no clear meltinglike transition. 

\end{abstract}


\maketitle

\section{\label{sec:intro}Introduction}

Small particles and clusters can have markedly different physical and chemical properties
from those of the bulk material because of the enhanced ratio of surface to volume and
quantum confinement effects.\cite{haberland}  For example, the structure and atomic
coordination, as well as electronic and magnetic properties, of small particles can all
show new features.\cite{jena}  In addition, the thermodynamic and melting behavior of
clusters can have peculiar properties.  For instance, small clusters of Sn and Ga were
found to undergo a meltinglike transition at temperatures higher than the melting point of
the corresponding bulk material,\cite{jarrold-sn,jarrold-ga} contradicting the standard
paradigm that a small particle should melt at a lower temperature than the bulk because
of the effect of the surface.  This behavior was explained in terms of a difference in the
nature of the bonding between cluster and bulk, with small clusters of Sn and Ga both
having a highly covalent character.\cite{sab-sn,ga-prl}  Other experiments showed that
small clusters of Na with from 50 to 300 atoms did undergo a meltinglike transition at
temperatures lower than the bulk melting point, but at temperatures that varied
irregularly with the size of the cluster, implying a competition between geometric and
quantum effects.%
\cite{schmidt-a,schmidt-b,aguado-11,hock-09,zamith-10,*zamith-11}

Experimental\cite{schwoebel-88,schwoebel-89,roy-94,wang-97} and
theoretical\cite{schwoebel-89,roy-94,wright-90,zhuang-02} work has also shown that the
properties of clusters supported on a surface can be modified by the interaction with the
surface.  However, there is a lack of work on the thermodynamic properties and melting
behavior of supported clusters using first-principles, density-functional-based electronic
structure methods.

A recent motivation for studying theoretically the structural and thermal properties of
supported metallic clusters is provided by the cluster-catalyzed growth process of carbon
nanotubes (CNTs).\cite{cnano,puretzky-00}  In this process, the CNT grows from a small
cluster, typically of a transition metal (or its oxide), supported on a substrate.  It is
highly desirable to be able to produce large quantities of high-quality CNTs with
controlled properties,\cite{cnano} but many aspects of the growth process are poorly
understood.  For instance, the efficiency of the growth process can depend markedly on the
particular combination of substrate and cluster material chosen, as well as on other
experimental parameters such as the temperature and pressure.  Much recent theoretical
work has been devoted to the simulation of the cluster-catalyzed CNT growth
process.\cite{ding-04a,maru-04,ding-06b,shibuta-07a,harut-07,jiang-07}

In this context, Ding \textit{et al.}\cite{ding-06b} and Shibuta and
Maruyama\cite{shibuta-07a} considered a simplified model in which the surface was replaced
by an idealized smooth plane interacting with the cluster via a potential of Lennard-Jones
type.  Treating the cluster-surface interaction (Lennard-Jones well depth) as a variable
parameter, they showed that the melting temperature of clusters with several hundred atoms
increased monotonically as the strength of the cluster-surface interaction increased.
Using a similar model, Sarkar and Blundell\cite{sarkar-09} considered smaller clusters of
size 55 atoms and showed that there were detailed changes in the structure of the cluster
as the cluster-surface interaction varied, accompanied by steps in the melting
temperatures.  Also, Jiang \textit{et al.},\cite{jiang-07} in a study of the
thermodynamics of supported Fe-C clusters, have employed a smooth plane interacting with
the cluster via an effective Morse potential, the parameters of which were fitted to
\textit{ab initio} calculations.

Now, the above-mentioned calculations\cite{ding-06b,shibuta-07a,sarkar-09,jiang-07} all
use classical molecular dynamics (MD)\cite{allen} with parametric interatomic potentials
to simulate the thermodynamic properties of the supported clusters.  In this work, we
retain the simplified model of the surface, but employ instead an \textit{ab initio}
treatment of the intracluster interactions, within density functional theory
(DFT),\cite{ksgen} directly in both the thermodynamic simulations and the global
optimization procedure used to determine the lowest-energy cluster structure at 0~K.  The
cluster-surface interaction is treated as an effective classical external potential acting
on the ions of the cluster.  While DFT-based thermodynamic simulations are orders of
magnitude more expensive than classical MD simulations, they are nevertheless feasible and
have been used with success in the past, for example, to explain the peculiar melting
behavior of free (unsupported) Sn and Ga clusters.\cite{sab-sn,ga-prl}  Also, first
principles determinations of the melting temperature of small Na clusters were made in
Refs.~\onlinecite{chacko-05} and \onlinecite{lee-05}.  To illustrate our general approach,
in this work we revisit Na clusters, with the aim of showing that the model is capable of
accounting for subtle quantum effects in the metallic bonding.

The plan of the paper is as follows. In the next section we describe in some detail our
model and calculation procedures.  Then, in Sec.~\ref{sec:N4-22monolayer}, we consider the
particular case of a cluster-surface interaction that is sufficiently strong that the
clusters collapse into a monolayer, or two-dimensional (2D), structure. This gives a particularly illuminating example of the differences between a classical and a quantum treatment of the valence electron gas. For this case, we
determine the lowest-energy structures for sizes $4 \le N \le 22$.  In order to bring out
the quantum effects in the metallic bonding, we also perform calculations for comparison
with a classical interatomic potential.  In addition to revealing a competition between
geometric and quantum effects, in this section we also explore the properties of `2D
metallic clusters' and consider evidence for 2D (rather than 3D) quantum effects.  Next,
in Sec.~\ref{sec:N20thermo}, we consider in detail the thermodynamics and melting behavior of a supported Na$_{20}$ cluster within the model for three values of the cluster-surface interaction strength, once again bringing out the similarities and
differences with the same simulations performed with classical interatomic potentials.  We also explore the dimensional effect on the melting behavior when the clusters are predominantly 2D.  The conclusions are given in Sec.~\ref{sec:concs}.

\section{\label{sec:method}Model and methodology}

A simple system displaying the main physical properties of metallic clusters is provided
by Na,\cite{haberland} and we shall consider Na clusters throughout.  Our main approach
involves a DFT-based Kohn-Sham (KS) description\cite{ksgen} of the metallic cluster.  We
employ either the local-density approximation (LDA) or the generalized-gradient
approximation (GGA) with Vanderbilt's ultrasoft pseudopotential,\cite{vanderbilt} as
implemented in the \textsc{vasp} package.\cite{vasp,*kresse-96}  These approaches will be
denoted KS-LDA and KS-GGA, respectively, in the following.  In parts of the work, we also
use a simplified (and computationally faster) real-space KS method in the LDA
incorporating a soft, phenomenological, local pseudopotential of the form described by
Blaise \textit{et al.}\cite{blaise-97}  This approximate KS method (henceforth referred to
as `KS-soft') has been used with success to describe the fragmentation of charged Na
clusters\cite{blaise-01} and the melting of free (unsupported) Na
clusters.\cite{vichare-01}  The KS formalism yields an expression\cite{ksgen} for the
internal energy $E_{\text{clus}}$ of the free cluster (that is, not yet taking into
account explicitly the interaction with the surface).

For comparison, we also describe the metallic bonding within a Na cluster by a classical
many-body Gupta potential derived within the second-moment approximation
(SMA).\cite{gupta,rosato-89,cleri-93}  The internal energy of a free (unsupported) cluster
in this approach is given by
\begin{eqnarray} E_{\text{clus}} &=& A\sum\limits_i {\sum\limits_{j \ne i} {\exp \left[ { -
p\left( {\frac{{R_{ij} }}{{r_0 }} - 1} \right)} \right]} }   \nonumber\\ && - \xi
\sum\limits_i {\left\{ {\sum\limits_{j \ne i} {\exp \left[ { - 2q\left( {\frac{{R_{ij}
}}{{r_0 }} - 1} \right)} \right]} } \right\}^{1/2} }, \label{eq:gupta} \end{eqnarray}
where $R_{ij}$ is the distance between ions $i$ and $j$.  The first term in
Eq.~(\ref{eq:gupta}) is a repulsive potential of Born-Mayer form, and the second term is a
cohesive energy.  We take the parameters for Na from the work of Li \textit{et
al.}:\cite{li-98} $\xi=21.398$ mRy, $A=1.1727$ mRy, $p=10.13$, $q=1.30$, and
$r_0=6.99\,a_0$, where $a_0$ is the Bohr radius.  These authors fitted the parameters to a
database of total energies as a function of lattice constant (for both fcc and bcc
lattices) obtained from calculations in the LDA for bulk Na. The potential was then
checked by using it to predict bulk properties such as the equilibrium lattice constant
and bulk modulus. For instance, the bulk melting temperature $T_m$ predicted by the model
was found to be 333~K, compared to the experimental value of 371~K.

Following Ding \textit{et al.}\cite{ding-06b} (and the general approach of Shibuta and
Maruyama\cite{shibuta-07a}), the surface is modeled as an idealized smooth plane that
interacts with the cluster via a Lennard-Jones 9/3 potential, yielding an interaction
energy
\begin{equation}
E_{{\text{int}}} = \varepsilon \frac{{3\sqrt 3 }}
{2}\sum\limits_i {\left[ {\left( {\frac{\sigma }
{{Z_i }}} \right)^9  - \left( {\frac{\sigma }
{{Z_i }}} \right)^3 } \right]} 
\,,
\label{eq:lj93}
\end{equation}
where $Z_i$ is the coordinate of ion $i$ perpendicular to the surface.  This model for the
surface is used with both the SMA (\ref{eq:gupta}) and the KS treatments of the
intracluster interactions used to determine $E_{\text{clus}}$. In this way, the cluster
is constrained to lie in the vicinity of the minimum of the Lennard-Jones well, which is
located roughly a distance $\sigma$ above the $Z=0$ plane.  We choose the parameter
$\sigma = 0.3$~nm.  As was done in Refs.~\onlinecite{ding-06b} and
\onlinecite{shibuta-07a}, the Lennard-Jones well depth $\varepsilon$ may be treated as a
variable parameter describing the overall strength of the cluster-surface interaction.
The total energy of the cluster plus surface is 
\begin{equation} E_{{\text{tot}}} =
E_{\text{clus}} + E_{{\text{int}}} \,.
\label{eq:etot} 
\end{equation}

Using these models of a supported metallic cluster, we carry out a global optimization
procedure to search for the structure that minimizes $E_{{\text{tot}}}$, and perform
statistical simulations to extract thermodynamic averages and study the melting behavior
of the cluster.  With both the SMA potential (\ref{eq:gupta}) and the KS-soft
method,\cite{blaise-01,vichare-01} we determine the optimum structure at 0~K by means of a
basin-hopping algorithm.\cite{li-87} This algorithm involves a Monte-Carlo (MC) process,
combined with optimization to the nearest local minimum at each MC step, to explore the
minima on the potential-energy surface.  With the SMA potential, we use 10$^5$--10$^6$
hops (local minimization steps).  The basin-hopping procedure generates a sampling of
excited isomers as well. We store the first (i.e., lowest-energy) six excited isomers
found with the SMA potential in a library of structures, together with four more
higher-energy isomers chosen randomly from the complete set of local minima generated.

Now, the KS model is orders of magnitude more expensive computationally than the SMA
model, and to find the optimum cluster structure within the KS approach we proceed as
follows.  Using the library of structures generated with the SMA potential as seeds
(initial structures) for the MC process, we perform 50--100 basin hops within the KS-soft
model for each seed structure.  Thus, we perform in total 500--1000 basin hops for each
cluster size $N$, the smaller number applying to the larger clusters that we consider
having up to $N=22$ atoms.  For the smaller cluster sizes $N<10$, we find that on doubling
the total number of basin hops, we do not observe a change in the lowest-energy structure
found.  Note that a basin-hopping process based directly on the KS model, as we perform here, often yields structures that were not present at all among the isomers of the SMA model (see Sec.~\ref{sec:N20thermo} for examples), and therefore we do not simply relax large numbers of isomers found within the SMA model.
To speed up the basin-hopping process, we use a relatively small simulation box
for the KS solution of side 45$\,a_0$.  For the larger sizes $N > 8$ considered, there are
usually many isomers with closely-spaced total energies (see Sec.~\ref{sec:N20thermo}),
and in some cases the ordering of these isomers can be modified by confinement effects for
a simulation box this small.  Therefore, we perform a second step in which the ten
lowest-energy structures found from the basin-hopping step are relaxed to a precise energy
tolerance ($\delta E_{\text{tot}} \sim 10^{-8}$~Ha) using a larger simulation box of side
90$\,a_0$ (but the same real-space grid spacing).  At this stage, in some instances (as
described in the text) we also relax the structures found within the KS-LDA or KS-GGA
treatments.\cite{vasp,*kresse-96}

Finally, we perform thermodynamic simulations for Na$_{20}$ following the general
procedures described in our earlier work,\cite{vichare-01} using both the SMA and KS-LDA
models of the cluster.  With the SMA model, we perform of order 30 constant-total-energy
(microcanonical) MD simulations distributed over a range of kinetic temperatures from 30~K
to around 750~K, with each simulation of order 1~ns (so that the total simulation time is
around 30~ns per cluster).  For kinetic temperatures higher than about 700~K, the Na
clusters tend to evaporate on this time scale.  We start at low kinetic temperature using
the optimum structure found above, and increase the kinetic temperature gradually from run
to run, using the coordinates and rescaled velocities at the end of one run to provide the
initial condition for the next run.  Thus, in effect we slowly heat the cluster.  The
kinetic temperatures chosen for the MD simulations are more closely spaced in the range
200~K to 450~K, where the cluster meltinglike process tends to occur.  After this, we
perform a multiple histogram fit to the overlapping histograms of potential energy from
the various simulations to extract the classical ionic density of states $\Omega(E)$.
With $\Omega(E)$ in hand, one can now evaluate thermodynamic averages such as the ionic
specific heat in a variety of ensembles, including the canonical
ensemble.\cite{vichare-01}

In the case of the KS-LDA cluster model, we also proceed via a slow-heating algorithm,
this time performing a sequence of isokinetic Born-Oppenheimer MD simulations\cite{payne}
at gradually increasing kinetic energies.  Because the KS-LDA approach is much more
expensive than SMA, we use a total simulation time per cluster that is somewhat smaller
than for the SMA model, of about 1~ns to 3~ns.  Finally, a canonical multiple histogram
fit is used to extract $\Omega(E)$.\cite{vichare-01}

\section{\label{sec:N4-22monolayer}Monolayer structures for $\protect\bm{N=4}$--$\protect\bm{22}$}

\begin{figure*}[tb]
\includegraphics[scale=0.85]{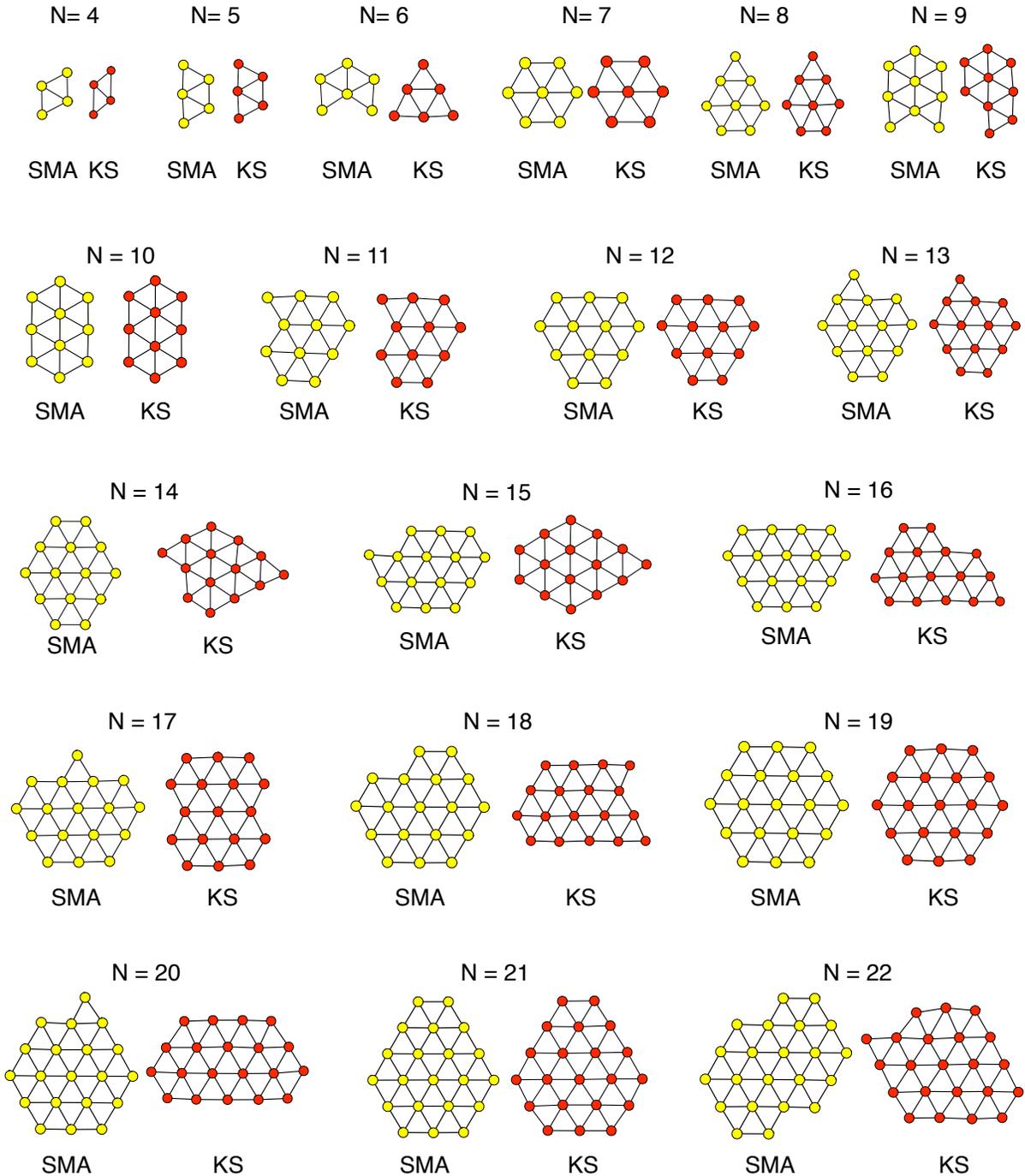}

\caption{ (Color online) Monolayer structures found for a cluster-surface interaction
strength $\varepsilon=0.5$\,eV using intracluster interactions from the second-moment
approximation (SMA), Eq.~(\ref{eq:gupta}), and the Kohn-Sham (KS) approach.}

\label{fig:N4-22monolayer}
\end{figure*}

\begin{figure}[tb]
\includegraphics[scale=0.5]{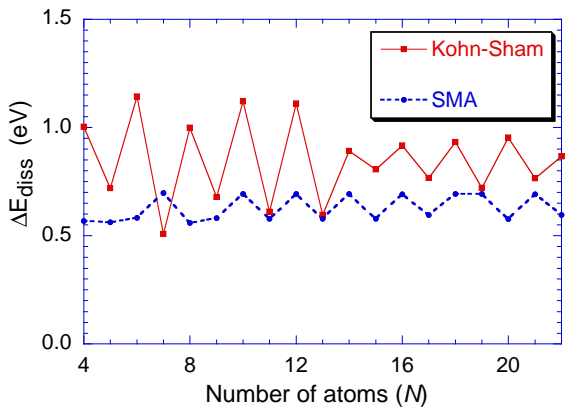}

\caption{ (Color online) Dissociation energies $\Delta E_{\text{diss}}$ of the supported
monolayer structures of Fig.~\protect\ref{fig:N4-22monolayer}, corrected for the
cluster-surface interaction strength $\varepsilon=0.5$\,eV.} 

\label{fig:Ediss}
\end{figure}

As we shall see in the next section, when the cluster-surface interaction $\varepsilon$
[see Eq.~(\ref{eq:lj93})] is increased, the cluster tends to become progressively flatter.
Useful insight into the role of quantum versus geometric effects can be gained by
considering a cluster-surface interaction sufficiently strong that the lowest-energy
structures are all monolayer.  For this purpose we have chosen $\varepsilon = 0.5$\,eV.
We have found optimum structures for both the SMA model of intracluster interactions,
Eq.~(\ref{eq:gupta}), and the KS-soft model\cite{blaise-01,vichare-01} using the
basin-hopping algorithm, as described in Sec.~\ref{sec:method}.  The final structures are
shown in Fig.~\ref{fig:N4-22monolayer}.  We also give the dissociation energies $\Delta
E_{\text{diss}}$ (the energy required to remove a single neutral atom from the cluster)
for these structures in Fig.~\ref{fig:Ediss}.  Noting that the atoms of a monolayer
structure all lie at the minimum $-\varepsilon$ of the Lennard-Jones well,
Eq.~(\ref{eq:lj93}), we have here defined
\begin{equation} \Delta E_{{\text{diss}}}(N)  = E_{{\text{tot}}} (N - 1) -
E_{{\text{tot}}} (N) - \varepsilon \,, \label{eq:Ediss} \end{equation}
correcting for the trivial contribution arising from the cluster-surface interaction.
Thus $\Delta E_{{\text{diss}}}(N)$ represents just the `internal' contribution to the
dissociation energy due to the intracluster interactions.

We see that the SMA structures in Fig.~\ref{fig:N4-22monolayer} form a geometrical packing
in a triangular lattice, favoring in particular hexagonal cluster structures.  Thus, for
$N=7$ and $N=19$ atoms the structures are two- and three-shell hexagonal arrangements,
respectively.  These are examples of compact, `closed-geometric-shell' structures in 2D.
Similarly, the structure for $N=10$ is a combination of two hexagonal structures.  The
structures either side, for example those at $N=18$ and $N=20$, are formed by adding or
subtracting one atom to or from the outside these closed-shell geometric structures.  In
fact, a striking property of the SMA structures in Fig.~\ref{fig:N4-22monolayer} is that
for all $4 \le N \le 21$ the structure for any size $N$ is obtained simply by adding one
atom to the outside of the structure for $N-1$.  This simple growth sequence is broken for
the first time at $N=22$.

The dissociation energy of the SMA structures (see Fig.~\ref{fig:Ediss}) mirrors this
growth sequence.  Thus, for the smaller sizes there are peaks at $N=7$ and 10, reflecting
the stability of these compact geometric structures.  There is then a sawtooth pattern in
$\Delta E_{\text{diss}}(N)$ for $N=10$--18 and $N=19$--22.  We can understand this
sawtooth variation in a very simple way by counting the number of nearest-neighbour bonds
in Fig.~\ref{fig:N4-22monolayer} broken in each dissociation.  For $N=10$--18, the atom to
be removed from an even-$N$ structure is connected by three bonds to the remainder of the
cluster, while that for an odd-$N$ structure is connected by only two.  In consequence,
the even-$N$ structure has a higher dissociation energy.  The sawtooth is interrupted at
$N=18$ and 19;  because of the geometrical properties of the growth sequence, the atom to
be dissociated has three bonds for both of these sizes.  The sawtooth pattern then resumes
after $N=19$, but this time it is the \textit{odd\/} sizes $N$ that have three bonds and
the greater dissociation energy.  Note also that it tends to be an atom on a corner that
is removed in a dissociation.  This happens because it is these atoms that have two or
three bonds to the rest of the cluster; an atom in the center of a complete side has four
bonds.

Turning to the KS structures in Fig.~\ref{fig:N4-22monolayer}, we note that there are some
similarities and some differences with the SMA structures.  Up to size $N=13$, the KS
structures also follow a simple growth pattern in which atoms are added successively to
the outside of each structure, although the structures for $N=6$ and 9 differ from their
SMA counterparts. For larger sizes $N > 13$, this simple growth sequence is broken.  Some
larger structures coincide with their SMA counterpart ($N=19$ and 21), but the others
generally prefer a more elongated form.  Also, the KS dissociation energies in
Fig.~\ref{fig:Ediss} form a consistent sawtooth pattern for all $4 \le N \le 22$.

These differences between the KS and SMA approaches are indicative of the role of quantum
effects in the system of valence electrons that forms the metallic bonding in the cluster.
The SMA potential favors a purely geometrical packing, and the dissociation energies are
sensitive to the number of nearest-neighbor bonds.  In metallic clusters, however, the
valence electrons display a fermionic `shell' structure---a finite-size quantum
effect---which is particularly pronounced at the smaller sizes considered
here.\cite{haberland,jena} A cluster with a closed fermionic shell is particularly stable,
in analogy to the noble gases in the periodic table. In the case of free (unsupported) Na
clusters, the closed shells coincide with those of the 3D simple harmonic oscillator
(SHO), namely, for $N=2$, 8, and 20 electrons.\cite{haberland,jena} Since in a Na cluster
each atom contributes one valence electron, these are also the (`magic') numbers of atoms
for which there is pronounced stability. For the monolayer structures discussed here, one
might expect that it is the closed shells of the 2D SHO that are appropriate: $N=2$, 6,
12, and 20.  These magic numbers occur in quasi-2D semiconductor quantum dots.\cite{jacak}

For a closed-shell system, the valence-electron gas is generally circular (in 2D) or spherical (in 3D), but between closed shells, the valence-electron gas tends to minimize its energy by
spontaneously deforming. This phenomenon is well-known in nuclear physics,\cite{bohrmottelson2} and has been shown to apply also to small, free 3D metallic clusters.\cite{koskinen-95}  For small cluster sizes, the deformation of the valence-electron gas can in turn drive the distribution of Na$^+$ ions away from circular symmetry, producing not just a distortion, but a new structure.  We see evidence for this phenomenon in Fig.~\ref{fig:N4-22monolayer} in the tendency of the KS model to yield elongated ground-state structures (relative to the SMA
structures) for $N > 12$, as is especially noticeable in the size range $N=14$--18. (The
case $N=20$, discussed in more detail in the next section, is an exception, because $N=20$
is expected to have closed fermionic shells, yet the optimum KS structure is still
somewhat elongated.)  We also see from Fig.~\ref{fig:N4-22monolayer} that the compact
geometric structure can still have the lower energy in particularly favorable cases, for
example, the three-shell hexagonal structure at $N=19$.  Thus, there is evidence of
competition between geometric and quantum effects in the KS structures.

The role of quantum effects is particularly striking in the dissociation energy,
Fig.~\ref{fig:Ediss}, where one finds a consistent sawtooth pattern in which the odd sizes
always have the smaller dissociation energy.  This effect, observed also in free metallic
clusters,\cite{haberland,jena} is a result of the pairing of spins: to a first
approximation, the spins for $N$ even form pairs (spin up and down) in each
single-particle energy level, yielding a more stable structure than for $N$ odd, which has
an unpaired final spin.  Note that while the SMA model also yielded a sawtooth pattern for
some sizes, this had a geometric origin and it could be either $N$ even or $N$ odd that
gave the smaller dissociation energy.  Also, the amplitude of the odd-even oscillations in
Fig.~\ref{fig:Ediss} tends to be greater for the KS model.

There may also be indications of 2D fermionic shell closures in the KS dissociation
energies in Fig.~\ref{fig:Ediss}.  It is striking that the amplitude of the odd-even
oscillations decreases abruptly for $N > 12$, which could be because a new fermionic shell
has been begun following a closed shell at $N=12$.  The evidence for a 2D fermionic shell
at $N=6$ is less clear, although the dissociation energy for the system with one
additional atom, $N=7$, is particularly low.  The evidence for a shell closure at $N=20$
seems unclear, however, either in the structures or in the dissociation energy.

\section{\label{sec:N20thermo}Thermodynamics and melting for $\protect\bm{N=20}$}

\begin{figure}[tb]
\includegraphics[scale=0.5]{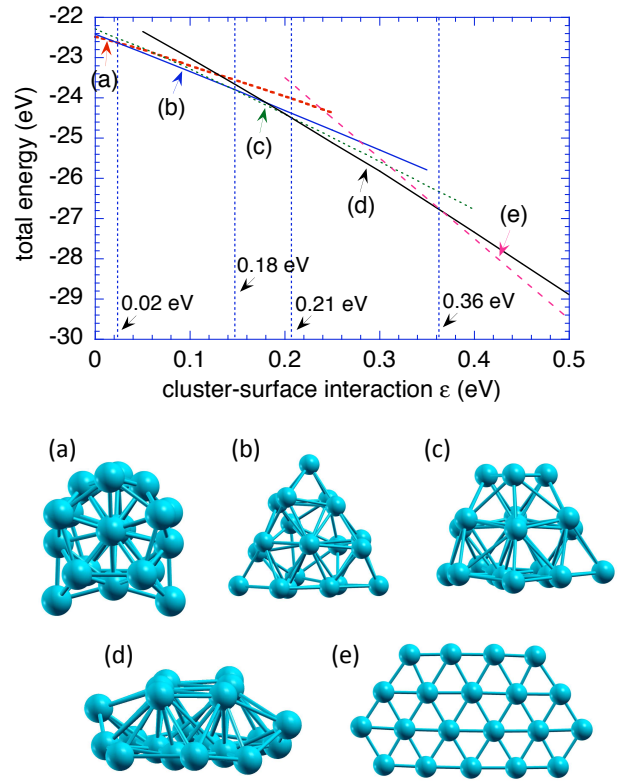}

\caption{ (Color online) Total energy of a Na$_{20}$ cluster on a surface (upper panel) as
a function of cluster-surface interaction strength $\varepsilon$ [see
Eq.~(\protect\ref{eq:lj93})] for the structures (a)--(e) shown in the lower panel.
Structure (a) (short-dashed line) is the lowest-energy structure for $0 \leq \varepsilon
\leq 0.02\,$eV; structure (b) (full line) for $0.02\,\text{eV} \leq \varepsilon \leq
0.18\,\text{eV}$; structure (c) (dotted line) for $0.18\,\text{eV} \leq \varepsilon \leq
0.21\,\text{eV}$; structure (d) (full line) for $0.21\,\text{eV} \leq \varepsilon \leq
0.36\,\text{eV}$; and structure (e) (dashed line) for $\varepsilon \geq 0.36\,\text{eV}$.}

\label{fig:crossover}
\end{figure}

In this section, we discuss in detail the structure and thermodynamics of a supported
Na$_{20}$ cluster within our model. Figure~\ref{fig:crossover} shows the lowest-energy
structures of Na$_{20}$ as a function of the cluster-surface interaction strength
$\varepsilon$ [see Eq.~(\ref{eq:lj93})].  The intracluster interactions in this figure are
treated within the KS-LDA approach. After determining candidate lowest-energy structures
at values of $\varepsilon$ from 0 to 0.5~eV in steps of 0.05~eV, we made a final series of
high-precision relaxations in order to plot the total energy (\ref{eq:etot}) of each
structure found as a function of $\varepsilon$ (see the upper panel of
Fig.~\ref{fig:crossover}).  The crossovers of these energy curves enable us to identify a
range of $\varepsilon$ values for which a particular structure is the lowest-energy
structure (that we have found).  In this way, we find a series of structural transitions
at particular values of $\varepsilon$, in which the cluster becomes progressively flatter
as $\varepsilon$ increases.  This behavior is similar to that found for small clusters of
Fe, Co, and Ni in Ref.~\onlinecite{sarkar-09} using classical methods with parametric
interatomic potentials.

The ground-state geometry of free Na$_{20}$ ($\varepsilon=0$),
Fig.~\ref{fig:crossover}(a), is quite spherical in shape. For small values of the
cluster-surface interaction strength $\varepsilon \approx 0.02$~eV, as the cluster-surface
interaction starts to become comparable to the internal interactions within the cluster,
the lowest-energy structure deforms from the spherical shape to a slightly elongated,
though still quite compact, form with the largest face aligned parallel to the surface
[Fig.~\ref{fig:crossover}(b)]. As $\varepsilon$ increases further, the lowest-energy
structure eventually flattens, first into a structure with three ionic layers
[Fig.~\ref{fig:crossover}(c)], and then into a two-layer structure
[Fig.~\ref{fig:crossover}(d)]. Finally, the structure becomes monolayer for $\varepsilon
\agt 0.36$~eV [Fig.~\ref{fig:crossover}(e)]. The flatter structures lower their energy by
increasing the contact area between the cluster and the surface for larger $\varepsilon$.

\begin{figure}[tb]
\includegraphics[scale=0.35]{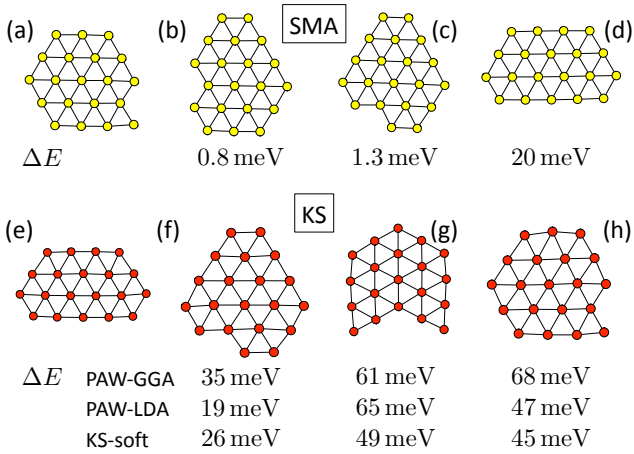}

\caption{(Color online) Ground-state (left column) and first three excited (right columns)
monolayer isomers of Na$_{20}$ found by the basin-hopping procedure for a cluster-surface
interaction strength $\varepsilon = 0.5\,$eV. Notation: SMA, second-moment approximation;
KS, Kohn-Sham; PAW, projected-augmented plane wave; GGA, generalized gradient
approximation; LDA, local density approximation; KS-soft, the simplified Kohn-Sham model
in the LDA with a soft, phenomenological pseudopotential (see text).  Excitation energies
$\Delta E$ are shown relative to the ground-state structure.}

\label{fig:na20_isomers}
\end{figure}

Let us consider the isomers in more detail in one case, that of the monolayer structure,
Fig.~\ref{fig:crossover}(e).  Figure~\ref{fig:na20_isomers} shows the ground and first
three excited isomers in the classical SMA potential and several KS models.  The SMA
isomers have been determined after $10^6$ basin hops.  For the KS isomers, we first
performed of order 500 basin hops within the KS-soft model (as discussed in
Sec.~\ref{sec:method}), retaining the lowest seven structures found.  A final
high-precision relaxation of these structures was made in the KS-LDA and KS-GGA approaches
[with a projected-augmented plane-wave (PAW) treatment of ionic cores, as implemented in
\textsc{vasp}\cite{vasp,*kresse-96}].

The three lowest-energy structures within the SMA [Figs.~\ref{fig:na20_isomers}(a)--(c)]
are nearly degenerate to within about 1~meV, reflecting the fact that these structures may
be regarded as surface rearrangements of each other.  Thus, it is possible to obtain the
ground-state SMA structure for $N=21$ in Fig.~\ref{fig:N4-22monolayer} by adding one atom
to the outer shell of each of these three structures.  

The situation is different with a KS treatment of intracluster interactions.  The KS
ground-state structure [Fig.~\ref{fig:na20_isomers}(e)] corresponds to an excited isomer
within SMA (lying at 20~meV), and is separated from the first excited isomer by about
20--35~meV.  (While there is some variation in the excitation energy of isomers according
to the KS model used, the ground-state structure is the same for each model.)  The first
excited isomer in the SMA [Fig.~\ref{fig:na20_isomers}(b)] appears to be disfavored in the
KS approach; it is not found by the basin-hopping procedure, and when we tried to relax it
within a KS approach it `slipped' to the geometry shown as the first excited isomer for KS
[Fig.~\ref{fig:na20_isomers}(f)].  Note that the first three excited isomers in the KS
model [Figs.~\ref{fig:na20_isomers}(f)--(h)] may be regarded as surface rearrangements of
each other and of the lowest three SMA isomers [Figs.~\ref{fig:na20_isomers}(a)--(c)].
However, the separation of these three isomers of KS is much greater than the 1~meV
separation found for the SMA structures.  These observations highlight the fact that the
energy of the metallic cluster in a quantum approach is sensitive, in a complicated way,
to the deformation and symmetry of the wave function, and not just to the geometric
packing of the ions or the number of nearest neighbors.

Note that excited structures such as (g) in Fig.~\ref{fig:na20_isomers}, or ground-state structures such as the KS structure for $N=16$ in Fig.~\ref{fig:N4-22monolayer}, are not found among the structures given by the SMA basin-hopping process (even after 10$^5$--10$^6$ hops), but become common in a KS basin-hopping process.  This illustrates the importance of basing the basin-hopping process directly on the KS model in order to achieve a fully unbiased search for the ground-state structure, despite the high computational cost of such an approach.

\begin{figure}[tb]
\includegraphics[scale=0.5]{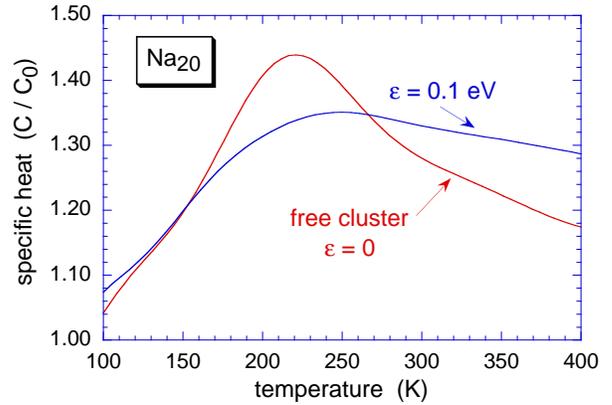}

\caption{ (Color online) Canonical ionic specific-heat capacities of Na$_{20}$, for
surface-cluster interaction strength [see Eq.~(\protect\ref{eq:lj93})]
$\varepsilon=0.1$~eV and for the free cluster ($\varepsilon=0$, taken from
Ref.~\protect\onlinecite{lee-05}), for KS descriptions of the intracluster interactions.
The specific heat is expressed as a multiple of its zero-temperature (classical) limit,
$C_0$.}

\label{fig:cv_eps0.1}
\end{figure}

Finally, we turn to the thermodynamic and melting behavior of the supported Na$_{20}$
cluster.  Figure~\ref{fig:cv_eps0.1} shows the canonical ionic specific-heat capacity of
the cluster for $\varepsilon = 0.1$~eV, extracted by an \textit{ab initio} MD simulation
and multiple-histogram fit using a KS-LDA description of the intracluster interactions
throughout (as described in Sec.~\ref{sec:method}).  For comparison, we also give the
specific-heat curve for a free Na$_{20}$ cluster ($\varepsilon = 0$) calculated by similar
methods, taken from Ref.~\onlinecite{lee-05}.  Inspection of the ionic trajectories shows
a meltinglike transition to occur for both $\varepsilon$ values, with the cluster passing
from a solidlike behavior at low temperature (vibration of ions around fixed points
combined with overall rotation) to a liquidlike behavior at high temperatures (diffusion
throughout the entire volume of the cluster).  The meltinglike transition is broad, with a
width of around 50--100~K and a `melting temperature' (conventionally corresponding to the
maximum on the specific-heat curve) around 220--240~K.  There is a significant change in
the specific-heat curve between the free and supported cluster: for $\varepsilon = 0.1$~eV
the curve is broader and the meltinglike transition correspondingly less well-defined, and
the melting temperature shifts to a slightly higher value.  This marked change in the
specific-heat curve is associated with a change in the lowest-energy structure at 0~K (see
Fig.~\ref{fig:crossover}).  

\begin{figure}[tb]
\includegraphics[scale=0.5]{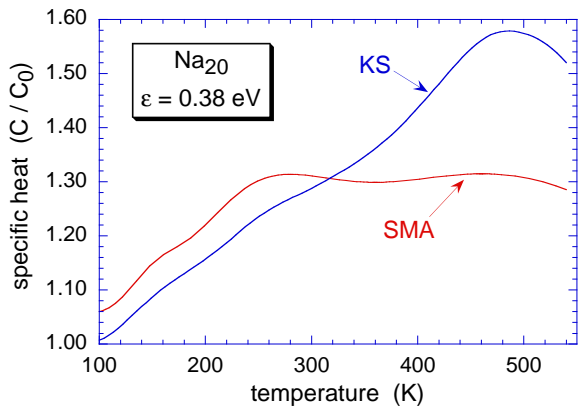}

\caption{ (Color online) Canonical ionic specific heat capacity of Na$_{20}$, with
surface-cluster interaction strength [see Eq.~(\protect\ref{eq:lj93})]
$\varepsilon=0.38$~eV, for SMA and KS descriptions of the intracluster interactions.  The
specific heat is expressed as a multiple of its zero-temperature (classical) limit,
$C_0$.}

\label{fig:cv_eps0.38}
\end{figure}

\begin{figure}[tb]
\includegraphics[scale=0.5]{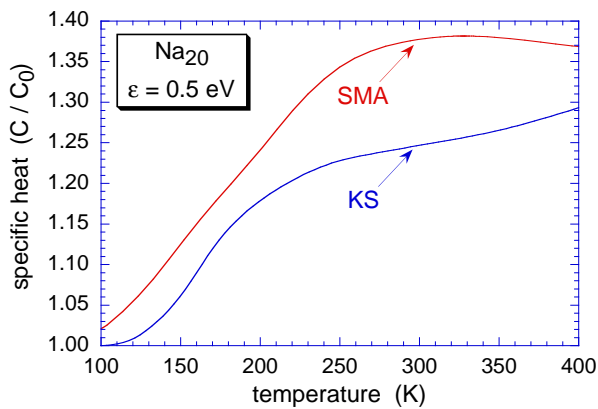}

\caption{ (Color online) Canonical ionic specific heat capacity of Na$_{20}$, with
surface-cluster interaction strength $\varepsilon=0.5$~eV.  See
Fig.~\protect\ref{fig:cv_eps0.38} for other notation.}

\label{fig:cv_eps0.5}
\end{figure}

Canonical ionic specific-heat curves are shown in Figs.~\ref{fig:cv_eps0.38} and
\ref{fig:cv_eps0.5} for $\varepsilon = 0.38$~eV and $\varepsilon = 0.5$~eV, respectively.
For both these values of $\varepsilon$, the lowest-energy structure at 0~K is a monolayer
structure (see Fig.~\ref{fig:crossover}), but the choice $\varepsilon=0.38$~eV for
Fig.~\ref{fig:cv_eps0.38} is very close to the critical value where the lowest-energy
structure becomes a two-layer structure.  This leads to an interesting difference in the
melting behavior.

For $\varepsilon=0.38$~eV, inspection of the ionic trajectories shows that, when the
cluster is liquidlike, ions frequently hop to the second layer and then back to the first,
usually reentering the first layer at a different position from where they left.  This
process occurs readily because there are energetically close isomers with a two-layer
structure.  We find a clear peak in the specific-heat capacity around 480~K, corresponding
to the temperature at which the ions first have sufficient energy (on average) to hop to
the second layer (that is, to overcome the barrier required to move to an energetically close two-layer structure).  For higher temperatures, the cluster is found to be in a liquidlike
state, and the mechanism of hopping to the second layer and back contributes significantly to the overall diffusion of the ions within the cluster.

On the other hand, for $\varepsilon=0.5$~eV, one finds that hops to the second layer are
highly suppressed.  At 300~K, for instance, the system spends less than 1\% of its time in
a two-layer configuration, and the ionic dynamics is thus constrained mostly to a plane
(2D).  At the lower temperatures $T \alt 300$~K, parts of the ground-state triangular lattice become distorted owing to the random movement of the ions, yielding distorted triangles or rectangles, for example.  The ionic motion becomes gradually more diffusive
as the temperature increases.  For temperatures above 400~K, one finds that the ions also begin to move increasingly in a vertical direction, toward the second layer.  The increasing phase space thus available to the ions means that the specific heat continues to rise beyond $T=400$~K, and there is no clear peak in the specific heat.  Around $T \agt 600$~K, the ions of the cluster tend to evaporate on the time scale of several hundred picoseconds used for the sampling runs.

We thus demonstrate clear dimensional effects in the melting of the clusters. When the ions are constrained to move mostly in 2D, as they are at $\varepsilon=0.5$~eV, the possibilities for diffusion are reduced compared to the 3D case (free Na clusters), and the onset of a liquidlike behavior is gradual, with no obvious peak in the specific-heat curve.  At intermediate values of $\varepsilon \approx 0.4$~eV, the motion is still predominantly 2D, but there are energetically close two-layer isomers; when the barrier to these configurations can be overcome, the sudden availability of additional phase space can give a peak in the specific-heat curve and a meltinglike transition.  Even in this case, however, we note that the observed melting temperature of $T_m \approx 480$~K is significantly \textit{above} the melting point of bulk Na, 371~K, contrary to the case of small free Na clusters,\cite{schmidt-a,schmidt-b} which melt at temperatures below that of the bulk (and contrary to the general paradigm for the melting of a small particle\cite{haberland}).

For comparison, we also give in Figs.~\ref{fig:cv_eps0.38} and \ref{fig:cv_eps0.5} the
specific heat calculated for SMA interatomic interactions.  Inspection of the ionic
trajectories shows that the mechanisms of melting are qualitatively similar to the
mechanisms in the KS model, but there are quantitative differences in the specific-heat
curves.  Thus, for $\varepsilon=0.38$~eV the meltinglike transition occurs around 260~K, a
lower temperature than observed with the KS model.  Similar differences in melting
temperatures were observed for free (unsupported) Na clusters in
Refs.~\onlinecite{chacko-05} and \onlinecite{lee-05}.  In these references it was found
that \textit{ab initio} DFT-based melting temperatures were in significantly better
agreement with experiment than those calculated with the SMA potential.

\section{\label{sec:concs}Conclusions}

We have presented a model for a supported metallic cluster that employs a first principles
DFT-based description of the cluster, but approximates the surface as an idealized smooth
plane interacting with the cluster via a parametric potential (here taken to be of
Lennard-Jones type).  Thermodynamic simulations, or global optimization methods such as
basin hopping, using this model are comparable in computational cost to similar methods
with the free clusters.  The results of this model were compared in detail with one in
which the intracluster interactions were described instead by a parametric interatomic
potential of SMA type, showing that important quantum effects were absent in the SMA-type
simulations for small metallic clusters.  The classical SMA model favors a purely
geometrical packing and is sensitive to the number of nearest neighbors, while a quantum
treatment of the valence electrons (as provided by a KS model) yielded significant
differences both in the lowest-energy structures and in energetic properties such as the
dissociation energy of the supported clusters. These differences could be ascribed in part
to fermionic shell closures, and we found some evidence that these shell closures could be
2D rather than 3D for the monolayer cluster structures that occur for large
cluster-substrate interactions.

We showed that the Kohn-Sham and Gupta models could yield significant quantitative differences in melting temperatures and caloric curves.  We also demonstrated clear dimensional effects in the melting of supported clusters by considering the case of a high cluster-surface interaction strength yielding a 2D monolayer structure.  For intermediate values of the cluster-surface interaction strength $\varepsilon \approx 0.4$~eV, the Na$_{20}$ cluster gave a clear meltinglike transition, with a peak in the specific-heat curve, but at an above-bulk melting temperature.  In this case, there were energetically close two-layer cluster structures.  At higher cluster-surface interaction strengths $\varepsilon \agt 0.5$~eV, where the ionic motion is constrained to be highly 2D, the onset of liquidlike behavior is gradual and there is no clearly definable `melting temperature.' Even for small values of the cluster-surface interaction strength, $\varepsilon \approx 0.1$~eV, the melting temperatures and caloric curves could be markedly different from those of the free cluster.

An obvious generalization of the present model would be to consider an atomistic surface,
and to extend the DFT treatment to the atoms of the surface.  Such an approach for
thermodynamic properties would be computationally very expensive, however, and the present
model is a compromise solution, useful for a class of systems.  It is likely to be realistic whenever the metallic bonding is not significantly disrupted by interaction with the substrate, which may be the case for an insulating substrate, for example.  Our approach might also be useful in
situations where the main interest is in the properties of the cluster away from the surface,
such as the interaction between the atoms of a carbon nanotube and the cluster in a
description of the nanotube growth process,\cite{ding-04a} where the nanotube grows from
the top of the supported cluster.

\begin{acknowledgments}

We gratefully acknowledge support from the European Commission through the EU-IndiaGrid2
network, \textit{Sustainable e-Infrastructures across Europe and India}, Contract
No.~RI-246698. SH would like to acknowledge an Indo-Swiss grant for financial support (No:
505 INT/SWISS/P-17/2009). We also thank C-DAC (Pune, India) for providing computer time
for some of the calculations. Some of the figures were generated by using VMD\cite{vmd}
and XCrySDen.\cite{xcrysden}

\end{acknowledgments}

%


\end{document}